\documentclass{emulateapj}
\usepackage{verbatim,graphicx,natbib}
\shortauthors{Cowan, Abbot \& Voigt}
\shorttitle{A False Positive for Glint on Unresolved Planets}
\begin{document}

\title{A False Positive for Ocean Glint on Exoplanets: the Latitude-Albedo Effect}
\author{Nicolas B. Cowan\altaffilmark{1,2}, Dorian S. Abbot\altaffilmark{3}, Aiko Voigt\altaffilmark{4}}
\altaffiltext{1}{Center for Interdisciplinary Exploration and Research in Astrophysics and Department of Physics \& Astronomy, Northwestern University, 2131 Tech Drive, Evanston, IL 60208, USA}
\altaffiltext{2}{CIERA Postdoctoral Fellow}
\altaffiltext{3}{Department of Geophysical Sciences, University of Chicago, 5734 South Ellis Avenue, Chicago, IL, 60637, USA}
\altaffiltext{4}{Max Planck Institute for Meteorology, Bundesstr. 53, D-20146, Hamburg, Germany}

\begin{abstract}
Identifying liquid water on the surface of planets is a high priority, as this traditionally defines habitability. One proposed signature of oceans is specular reflection (``glint''), which increases the apparent albedo of a planet at crescent phases. We post-process a global climate model of an Earth-like planet to simulate reflected lightcurves.  Significantly, we obtain glint-like phase variations even though we do not include specular reflection in our model.  This false positive is the product of two generic properties: 1) for modest obliquities, a planet's poles receive less orbit-averaged stellar flux than its equator, so the poles are more likely to be covered in highly reflective snow and ice, and 2) we show that reflected light from a modest-obliquity planet at crescent phases probes higher latitudes than at gibbous phases, therefore a planet's apparent albedo will naturally increase at crescent phase. We suggest that this ``latitude-albedo effect'' will operate even for large obliquities: in that case the equator receives less orbit-averaged flux than the poles, and the equator is preferentially sampled at crescent phase.  Using rotational and orbital color variations to map the surfaces of directly imaged planets and estimate their obliquity will therefore be a necessary pre-condition for properly interpreting their reflected phase variations.  The latitude-albedo effect is a particularly convincing glint false positive for zero-obliquity planets, and such worlds are not amenable to latitudinal mapping. This effect severely limits the utility of specular reflection for detecting oceans on exoplanets. 
\end{abstract}

\section{Introduction}
The traditional habitable zone (HZ) is defined in terms of surface liquid water \citep{Kasting_1993}. Three distinct methods have been proposed to search for liquids on the surface of a planet: 

\emph{Rotational color variability} \citep{Ford_2001, Cowan_2009, Kawahara_2010, Cowan_2011c, Kawahara_2011}: oceans are darker and have different colors than other surface types on Earth, so the time variations in color of a spatially unresolved planet can betray the presence of liquid water oceans.  This method relies on there being longitudinal inhomogeneities in the planet's surface composition.

\emph{Polarization} \citep{Zugger_2010, Zugger_2011}: oceans are smoother than other surface types and thus polarize light.  For idealized scenarios, the phase variations in polarization are significant, but the same authors found that in practice the effect of oceans is masked by Rayleigh scattering, clouds and aerosols. Observations of polarized Earthshine, however, imply that rotational variations in polarization may be useful in detecting oceans \citep{Sterzik_2012}.

\emph{Specular reflection} \citep{Williams_2008, Robinson_2010}: oceans are also able to specularly reflect light, especially at crescent phases. The signal-to-noise requirements for phase variations are not as stringent as for rotational variations since the integration times can be much longer: weeks instead of hours. However, \cite{Robinson_2010} showed that clouds not only mask underlying surfaces, but forward scattering by clouds mimics the glint signal at crescent phases, while atmospheric absorption and Rayleigh scattering mask the glint signature. They proposed using near infrared opacity windows to search for glint, but this would only be possible if the effects of clouds could be accurately modeled for exoplanets.  

Clearly, any method attempting to probe the planetary surface will be impeded by clouds, but the three techniques above have been shown to work in limited empirical and/or numerical experiments for planets like Earth, which is roughly half covered by clouds. Such tests are necessary but not sufficient.  While previous authors have noted that seasonal changes in snow, ice and cloud coverage lead to 15--20\% modulations in a planet's apparent albedo \citep[][]{Williams_2008, Oakley_2009, Robinson_2010}, they under-appreciated the importance of viewing geometry.  

In this Letter, we focus on the specular reflection method and show that the generic tendency of HZ planets to have snow and ice in their least-illuminated regions naturally leads to a false positive for ocean glint.\footnote{We demonstrate the latitude-albedo effect using a model with an ocean, but nearly dry planets develop similar latitudinal albedo profiles \citep{Abe_2011}.} This is because the viewing geometry changes throughout an orbit such that the least-illuminated latitudes are preferentially sampled at crescent phases. We call this the latitude-albedo effect.

\section{Methods}
\subsection{Global Climate Model}
We calculate reflected phase variations for a variety of viewing geometries using a simulation presented in
\cite{Voigt_2011} generated by the global 
climate model (GCM) ECHAM5/MPI-OM. It simulates the circulation
of the atmosphere and ocean and has interactive
sea-ice and clouds; a detailed description of the GCM
is given in \cite{Voigt_2011} and references
therein. ECHAM5/MPI-OM
includes a shortwave radiation model that calculates 
the top-of atmosphere incident ($F_{\downarrow}$) and reflected ($F_{\uparrow}$) shortwave radiative
flux at every time step and grid location of the model. 

The model has a 3.75$^\circ$ horizontal resolution and does not allow radiative transfer between columns. The shortwave 
scheme includes absorption and scattering.  Forward scattering from aerosols and cloud particles is incorporated by 
asymmetry factors and using the delta-Eddington approximation. The model accounts for the increased atmospheric path length at grazing angles of incidence, but surface reflection does not depend on solar zenith angle and there is no specular reflection in our GCM. 

This simulation uses  
a modern-Earth solar constant of 1367\,Wm$^{-2}$, pre-industrial
atmospheric greenhouse gases, a modern-Earth
orbit (23.5$^\circ$ obliquity, 1.7\% eccentricity) and continent configuration meant to
represent Earth 635~Ma before present. The
simulation is initiated from a climate state with polar ice caps and remains in
such a state throughout. It was run to equilibrium, then for an additional ten
years, with shortwave radiation outputs saved every two hours. The planet has a mean surface albedo of $0.244$ and is 57\% covered in liquid water, making it habitable by definition. Cloud cover is $0.654$ and the time-averaged planetary albedo is $0.351$, both slightly higher than for simulations of pre-industrial modern Earth.

\begin{figure}[htb]
\begin{center}
\includegraphics[width=84mm]{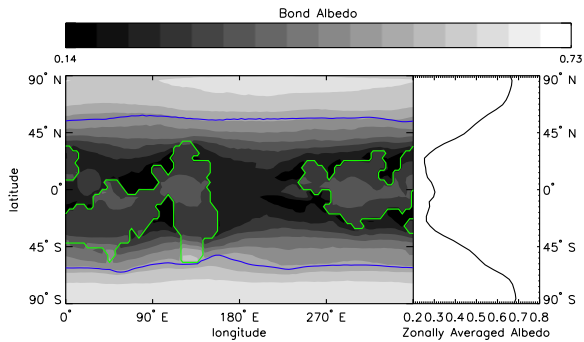} 
\end{center}
\caption{Ten-year averaged map of planetary albedo for the simulation. Green lines indicate the coast of the equatorial continents; blue lines show the annually averaged extent of sea-ice.  The right panel shows the zonally (longitudinally) averaged albedo profile.}
\label{average_map}
\end{figure}

Since we use top-of-atmosphere fluxes, all of the albedos in the remainder of this Letter are planetary albedos, as opposed to surface albedos. We define $A(\theta, \phi, t) \equiv F_{\uparrow}(\theta, \phi, t)/F_{\downarrow}(\theta, \phi, t)$, where $\theta$ and $\phi$ are planetary latitude and longitude, respectively. In Figure~\ref{average_map} we plot $\overline{A} = \overline{F_{\uparrow}}/ \overline{F_{\downarrow}}$, were overbars indicate time-averages.

We compute lightcurves with 2~hr time resolution, including rotational variations, as well as changes in cloud cover on diurnal, seasonal and inter-annual timescales.  For the purposes of this Letter we consider 24-hr integration times for the observations, which average over the rotational variations. 

\subsection{Bond Albedo}\label{bond_albedo}
We first consider the bolometric spherical albedo of the planet, or Bond albedo,
\begin{equation}
A_{B}(t) = \frac{\oint F_{\uparrow}(\theta, \phi, t) d\Omega}{\oint F_{\downarrow}(\theta, \phi, t) d\Omega},
\end{equation}
where $\oint d\Omega$ is the surface integral over the entire sphere. 
  
  The Bond albedo exhibits seasonal variations of 6\% (Figure~\ref{apparent_albedo_lat}). These are largely driven by the meridional gradient in albedo (right panel of Figure~\ref{average_map}) and are much smaller that the latitude--albedo effect.

\subsection{Disk-Integrated Reflected Light}
We now calculate light curves as they would appear to a distant observer, assuming diffuse (a.k.a. Lambertian) reflection. We completely neglect any specular reflection that could produce a glint spot, yet, as we show below, we still find that the apparent albedo exhibits the phase variations one would expect for glint.  

We define the normalized weight as $W(\theta,\phi,t) = V F_{\downarrow} /\oint V F_{\downarrow} d\Omega$, where $V(\theta, \phi, t)$ is the visibility of a given region of the planet for a given observer ($V$ equals one at the sub-observer point, drops as the cosine of the angle from the sub-observer point, and is zero on the far side of the planet). For a given viewing geometry, $W$ quantifies the sampling of different regions on the planet.

The apparent albedo is the disk-averaged albedo of the planet, weighted by illumination and visibility: 
\begin{equation}\label{apparent_albedo}
A^{*}(t) = \frac{\oint V F_{\uparrow} d\Omega}{\oint V F_{\downarrow} d\Omega} = \oint  W(\theta, \phi, t) A(\theta, \phi, t) d\Omega.  
\end{equation}

\section{Results}
\begin{figure*}[htb]
\begin{center}$
\begin{array}{cc}
\includegraphics[width=84mm]{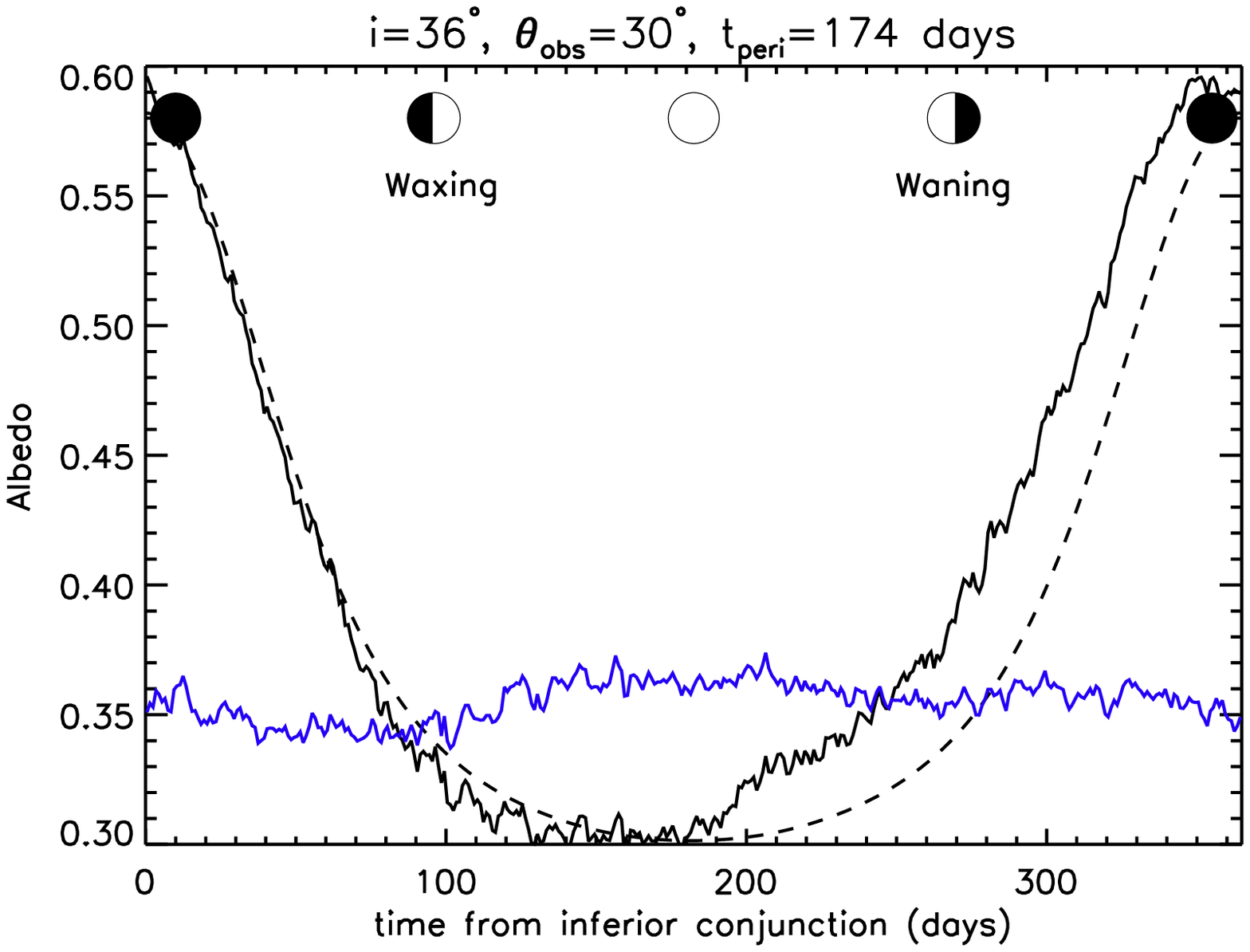} &\includegraphics[width=84mm]{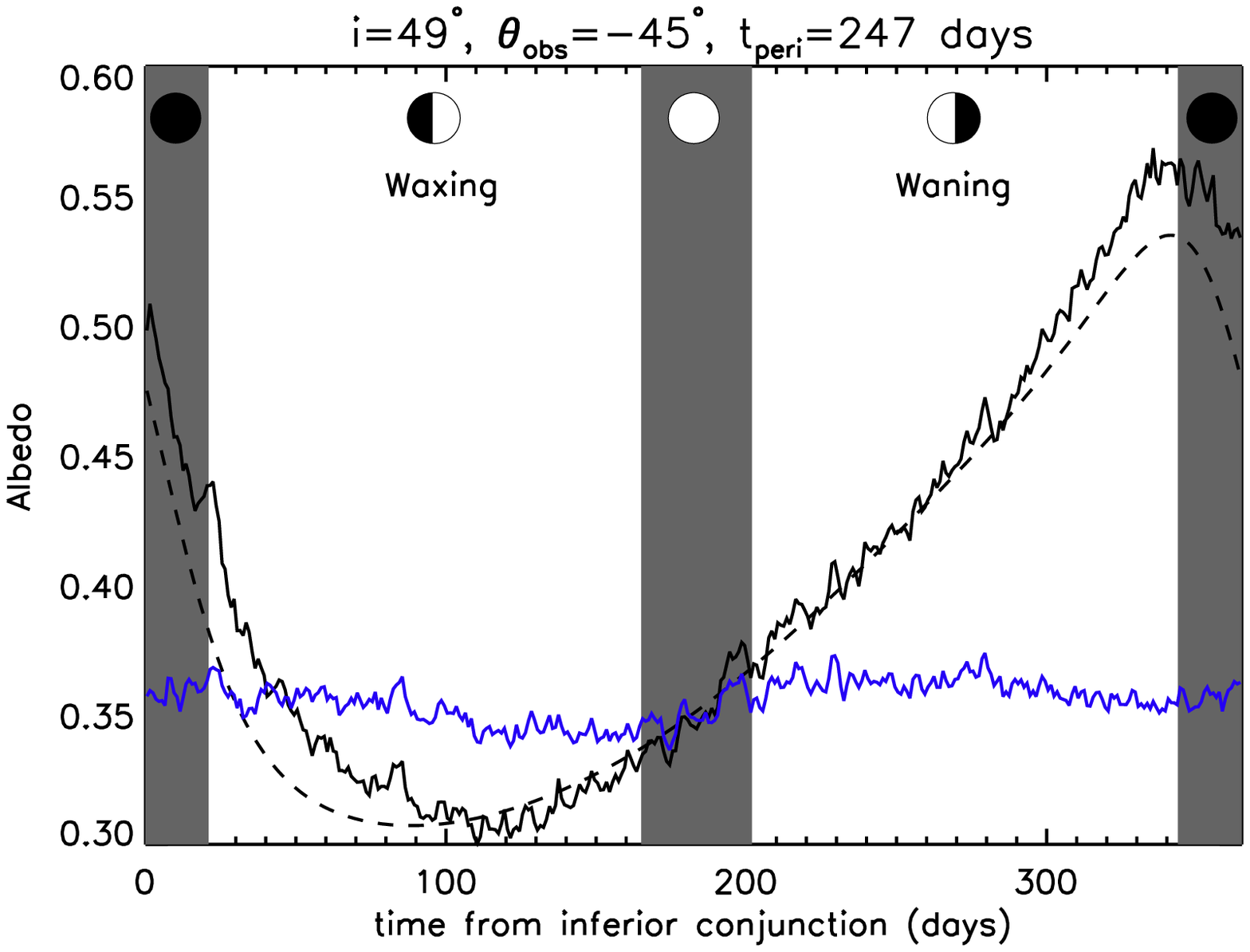}\\
 \includegraphics[width=84mm]{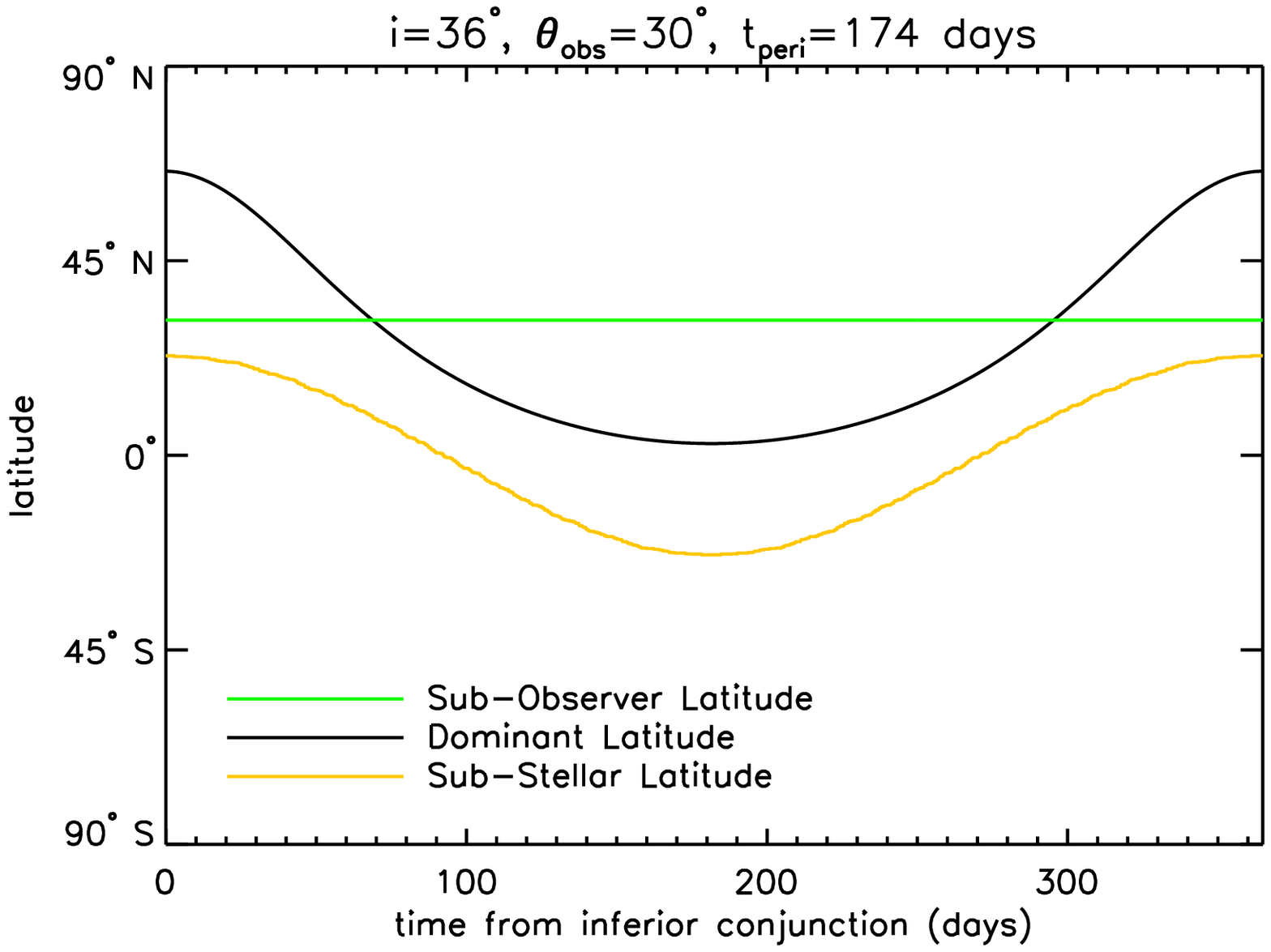}& \includegraphics[width=84mm]{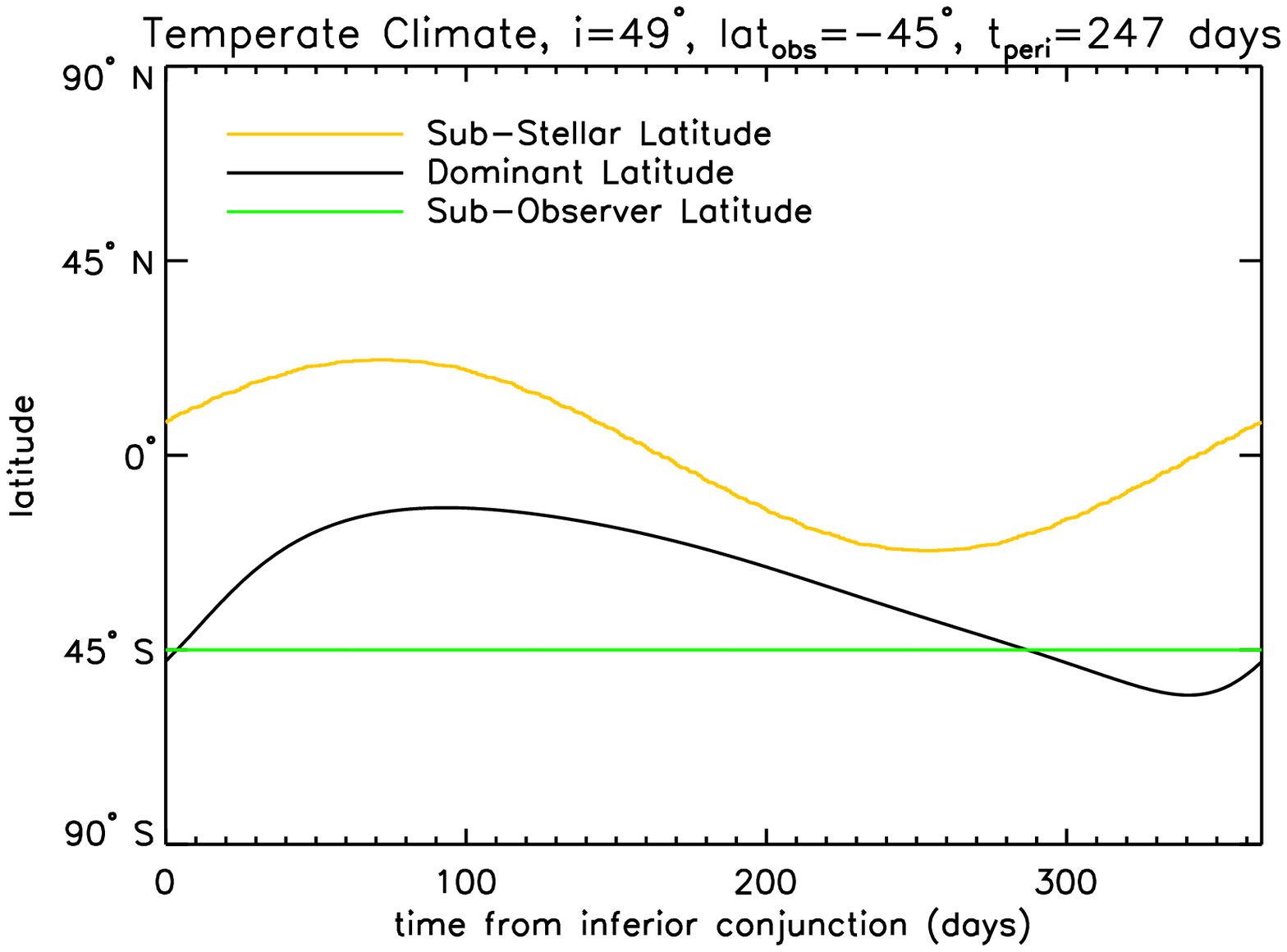}
\end{array}$ 
\end{center}
\caption{\emph{Top:} Phase variations in apparent albedo (solid black) compared to Bond albedo (blue) for two viewing geometries. The dashed black lines show the apparent albedo for a static, time-averaged map. The left and right edges of the plot correspond to crescent phases. The grey regions in the right panel denote the phases that will not be accessible because the planet will be inside the inner working angle (IWA) of the telescope; the geometry shown at left is sufficiently face-on that the planet is never within the IWA. \emph{Bottom:} Phase variations in the dominant latitude, $\theta_{\rm dom}$, for the same two viewing geometries. The orbital inclination, $i$, is the angle between the planet's orbital angular momentum and the line of sight ($i=0^\circ$ for a face-on orbit, $i=90^\circ$ is edge-on); the sub-observer latitude is denoted by $\theta_{\rm obs}$. The time of periastron, $t_{\rm peri}$, determines the orientation of the (slightly elliptical) orbit with respect to the line of sight. Northern winter solstice nearly coincides with periastron, as on present-day Earth.}
\label{apparent_albedo_lat}
\end{figure*}

In the top panels of Figure~\ref{apparent_albedo_lat} we show the phase variations in apparent albedo for two generic observing geometries (solid black lines). We gray-out the phases for which the planet would be inside the inner working angle (IWA) of a high-contrast imaging mission and therefore unobservable.  The IWA depends on the details of a mission's design and is a function of wavelength and distance. We gray-out phases within $45^{\circ}$ of full or new phase, corresponding to an IWA of 71~mas for a system at 10~pc.

The phase variations in apparent albedo  are shown with solid black lines in the top panels of Figure~\ref{apparent_albedo_lat}. Despite the fact that our model includes only diffuse reflection, the variations have approximately the same shape and amplitude as glint for an Earth-like geography and cloud cover \citep{Williams_2008, Robinson_2010}. 

For the left geometry of Figure~\ref{apparent_albedo_lat}, the planet/star flux ratio is $2.4\times10^{-10}$ and $0.6\times10^{-10}$ at superior and inferior conjunction, respectively. For an Earth-twin at 10~pc, a 10~m telescope with 5\% throughput could obtain 1\% photometry in a 100~nm optical band in 6~days of integration at inferior conjunction, when the planet is faintest \citep[following][]{Roberge_2012}.   

\subsection{Dominant Latitude} \label{theta_dom_sec}
The phase variations can be understood in terms of the dominant latitude and the planet's latitudinal albedo profile.  The dominant latitude is the latitude most sampled for a given viewing geometry \citep[][]{Cowan_2011c}:
\begin{equation}
\langle \theta \rangle = \oint W(\theta, \phi, t) \theta d\Omega.
\end{equation}

In the bottom panels of Figure~\ref{apparent_albedo_lat} we show the sub-stellar, sub-observer, and dominant latitudes for the same two geometries used in the top panels.  Significantly, the dominant latitude may be poleward of both the sub-stellar and sub-observer latitudes, but this only occurs at crescent phases (near inferior conjunction).  
 
The latitudinal albedo profile is shown in the right panel of Figure~\ref{average_map}. Albedo is lowest in the tropics, so the planet's apparent albedo will tend to be greater than its Bond albedo, because the dominant latitude is typically poleward of the sub-stellar latitude. Furthermore, variations in the dominant latitude over the course of a year couple with the latitude dependence of albedo to generate phase variations in the planet's apparent albedo (the latitude-albedo effect). 

We estimate the magnitude of the latitude-albedo effect by replacing the actual time-variable albedo of the planet with its ten-year average (Figure~\ref{average_map}) when calculating apparent albedo (Eq.~\ref{apparent_albedo}). Since the 10-year albedo map averages over diurnal and seasonal variations in albedo, this exercise isolates the latitude-albedo effect. The approximate albedo variations are shown by the dashed black lines in the top panels of Figure~\ref{apparent_albedo_lat}.  

Clearly the latitude dependance of albedo is the dominant factor enhancing the apparent albedo at crescent phase in our simulation. The discrepancies between the solid and dashed black lines are due to second-order effects. For example, seasonal ice and snow increase the apparent albedo during Boreal winter and spring (top-left panel of Figure~\ref{apparent_albedo_lat}), while the dependence of albedo on solar zenith angle provides an additional enhancement of apparent albedo at crescent phases (top-right panel of Figure~\ref{apparent_albedo_lat}). At large zenith angles, GCM pixels with clouds overlying dark surfaces (water or land) have enhanced albedos because the longer path-length through the atmosphere leads to more scattering.  This increases the brightness of the day--night terminator at low latitudes. Diurnal cycles in cloud coverage could also cause reflected phase variations \citep[for Earth these patterns are weak;][]{Hartmann_1991}.

\subsection{Statistical Analysis of Geometry}
The latitude-albedo effect is a confounding factor for the photometric detection of specular reflection. To be a convincing false-positive, however, the albedo must not only increase at crescent phases (true of both geometries in Figure~\ref{apparent_albedo}), but must also be symmetrical at waxing and waning phases (true for the left geometry, but not the right). The critical variable affecting these observables is the orientation of the planet's axial tilt with respect to the observer: the obliquity phase. 

We perform a statistical analysis of inclinations, obliquities, and obliquity phases to determine which viewing geometries might be convincing false positives.  We cannot predict how snow and ice coverage will vary as a function of obliquity without rerunning computationally expensive climate simulations, but we track dominant latitude as a function of orbital position for each geometry.  In particular, we consider $\langle|\theta|\rangle = \oint W |\theta| d\Omega$ since we only care about distance from the equator, not whether a region is in the Northern or Southern hemisphere.  

\begin{figure}[htb]
\begin{center}
\includegraphics[width=84mm]{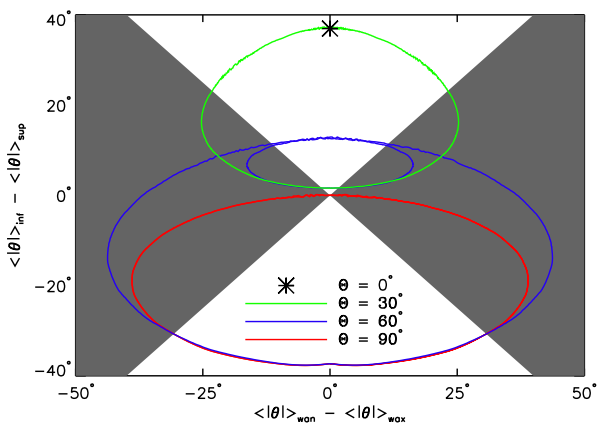}
\end{center}
\caption{Excursions in dominant latitude, $\langle|\theta|\rangle$, for planets with $60^\circ$ orbital inclination.  Each color shows the locus for a single obliquity, $\Theta$, and 1000 different obliquity phases; the black asterisk shows the zero-obliquity case. Positive amplitudes (top of plot) indicate that the regions probed at crescent phase are poleward of those sampled at gibbous phase; this is always the case for planets with modest obliquities, $\Theta=0^{\circ}$ and $30^\circ$. Negative amplitudes (bottom of plot) indicate that the regions sampled at crescent phases are equatorward of those probed at gibbous phases; this is always the case for $\Theta=90^{\circ}$.  Non-zero asymmetry (left and right edges of plot) indicate that the latitudes sampled at waning phases are different from those sampled at waxing phases.  The white regions correspond to $\langle|\theta|\rangle$-variations that are liable to be convincing glint false positives: the change at crescent phases will be greater than the asymmetry between waxing and waning phases. The fraction of planets falling in this ``dangerously convincing'' area is 100\%, 59\%, 30\% and 29\% for obliquities of $0^\circ$,  $30^\circ$,  $60^\circ$,  and $90^\circ$, respectively.}
\label{master_inclination}
\end{figure}

The most important quantities are 1) the amplitude of the excursions in dominant latitude: $\langle|\theta|\rangle$ at inferior conjunction minus that at superior conjunction (this underestimates the amplitude for asymmetric excursions), and 2) the asymmetry in $\langle|\theta|\rangle$ about superior conjunction, which we estimate as $\langle|\theta|\rangle$ at waning quarter minus $\langle|\theta|\rangle$ at waxing quarter.  In Figure~\ref{master_inclination} we adopt the most likely orbital inclination, 60$^{\circ}$, and plot the amplitude of dominant latitude excursions versus the asymmetry of those excursions.

Recall that for obliquities less than $53.9^\circ$ the equator receives more orbit-averaged insolation than the poles, while at greater obliquities the poles receive more flux than the equator, albeit with large seasonal variations. In nearly all cases shown in Figure~\ref{master_inclination}, the regions probed at crescent phase receive less orbit-averaged insolation than those probed at gibbous phases, which will lead to larger apparent albedos at crescent phase if these regions have year-round snow.  The exception is $\Theta=60^\circ$, for which the latitudes sampled at crescent phase may be poleward \emph{or} equatorward of those probed at gibbous phases; but a planet with such an obliquity has relatively uniform orbit-averaged insolation as a function of latitude and may not have year-round ``cold'' regions \citep[but see][]{Abe_2011}.  

The glint false positive is most extreme and pernicious for planets with zero obliquity because the dominant latitude at crescent phases is nearly 40$^\circ$ poleward of that at gibbous phases, and the excursions are perfectly symmetrical about superior conjunction (black asterisk at top of Figure~\ref{master_inclination}). For planets with non-zero obliquity, the latitude-albedo effect best mimics the glint signature (no asymmetry) when the orbital/viewing geometry is such that solstices coincide with conjunctions. 

The same effect occurs for edge-on inclinations (not shown), but in that case the dominant latitude only increases significantly at extreme crescent phases (e.g., within $\sim6^\circ$ of inferior conjunction, for $i=85^\circ$). 

\section{Discussion and Conclusions}
The latitude-albedo effect operate if the regions of the planet receiving the least orbit-averaged flux have the greatest albedo due to inhomogeneous surface or cloud cover.  

\cite{Williams_2003} simulated the climate of high-obliquity Earth analogs and did not find year-round equatorial snow, but they speculated that such coverage would occur for lower CO$_2$ levels, or higher-elevation continents.  \cite{Abe_2011} found that planets with no large surface liquid water reservoirs store their condensibles at their coldest latitudes, regardless of obliquity.   

Even if no surface snow/ice exists on a planet, or if the surface is obscured by a thick atmosphere, the latitude-albedo effect will still operate if the coldest regions of a planet are most cloudy. This is the case in our simulation, but may not be in general. Meridional gradients in cloudiness could be important for planets with surface temperatures too hot for snow and ice, due to low albedo, high insolation, high concentrations of greenhouse gases, or thicker atmospheres ---in short, planets at the inner edge of the HZ.  
    
If a planet's apparent albedo increases at crescent phases, there are two fundamentally different explanations: 1) the scattering
phase function of the planet is forward peaked due to, for example, atmospheric Rayleigh scattering, Mie scattering from clouds, or specular
reflection from water,  or 2) there are more scatterers in the scene, either intrinsically, as with the latitude--albedo effect, or because of path length effects.

If a planet's apparent albedo exceeds unity, then ---by definition--- it must be scattering non-diffusively, but this only occurs at or within the IWA for Earth-like models \citep{Robinson_2010}. More importantly, the albedo-radius degeneracy for directly imaged planets makes it impossible to put albedo on an absolute scale.

Glint polarization fraction peaks at quadrature and cannot directly distinguish between sources of brightening at crescent phase, although polarimetry could potentially detect surface water at other phases \citep{Zugger_2010, Zugger_2011}.

\cite{Robinson_2010} noted that glint measurements should be performed at near-infrared (NIR) wavelengths in order to avoid Rayleigh scattering, and in opacity windows in order to avoid atmospheric absorption. The reflectance of the glint spot is essentially the same at NIR wavebands as in the optical, while snow becomes less reflective at these longer wavelengths. The optical--NIR color of snow depends sensitively on grain size, with larger grains being redder \citep{Warren_1982}. It is therefore possible that measurements of reflected phase variations at both optical and near-infrared wavelengths could help disentangle the effects of ocean glint from that of large-grained snow, provided that Rayleigh scattering does not obscure the surface at the shorter wavelength.

To summarize, detecting ocean glint for an Earth-like planet based solely on multiband phase variations is only possible if the effects of clouds and snow can be properly modeled for extrasolar planets. For the remainder of this Letter we assume that the snow/glint degeneracy will not be resolved, and examine how the latitude-albedo effect is likely to affect exoplanet characterization.

High contrast imaging missions will be able to monitor the apparent albedo of terrestrial planets in the HZ of nearby Sun-like stars. In the near-term, pairing the James Webb Space Telescope \citep[JWST;][]{Gardner_2006} with a starshade \citep[][]{Soummer_2010} might achieve the same goal.

If a directly imaged terrestrial planet is observed for only half an orbit and its apparent albedo is greater at crescent than at gibbous phase, no conclusion can be drawn. If the observed albedo variations are monitored for a full orbit and are symmetrical, it will mean a) the planet has zero obliquity, b) the planet's solstices coincide with conjunctions, or c) the latitude-albedo effect is unimportant.  If, on the other hand, the lightcurve is strongly asymmetrical, one will suspect that the latitude-albedo effect is at play, but will be unable to correct for the effect, because of the unknown obliquity and obliquity phase of the planet. Note that asymmetry in apparent albedo may also be due to seasonal changes in the planet's intrinsic Bond albedo, but we found this to be a minor effect in our simulation. 

Fortunately, high signal-to-noise, high-cadence reflected light measurements of an imaged exoplanet can yield  precisely the variables one needs to correct for the latitude-albedo effect. This involves: 1. determining the rotation rate of the planet \citep[tested in simulations by][]{Palle_2008, Oakley_2009}, 2. making rotational albedo maps of the planet \citep[tested using EPOXI Earth observations by][]{Cowan_2009, Fujii_2011}, and 3. doing so at a variety of orbital positions \citep[tested in simulations by][]{Kawahara_2010, Kawahara_2011, Fujii_2012}.

It is therefore possible to properly interpret reflected light phase variations of imaged planets at crescent phases provided that one obtains high signal-to-noise, high-cadence measurements at gibbous phases \citep[for signal-to-noise considerations, see][]{Cowan_2009, Kawahara_2010}. 

Zero-obliquity planets are the worst case for three reasons: 1) variations in dominant latitude are large and symmetrical, 2) poles receive the least orbit-averaged flux, and are therefore most likely to harbor year-round snow, and 3) these planets are not amenable to latitudinal mapping, because the sub-stellar point is always equatorial.  Although planets with negligible obliquities will likely be a minority of directly-imaged worlds, they are the norm in the HZ of low-mass stars. The latitude-albedo effect will therefore be an important glint false positive for missions characterizing the reflected phase variations of temperate planets orbiting nearby M-Dwarfs.  
  
\acknowledgements
The idea for this paper arose during the ExoClimes 2012 meeting at the Aspen Center for Physics.  We acknowledge useful conversations with T. Robinson, constructive comments from our referee, as well as support from the German Research Foundation (DFG) program for the initiation and intensification of 
international collaboration.

\end{document}